# The Influence of Magnetic Fields on UHECR Propagation from Virgo A

**Oleh Kobzar**[1]
*Institute of Nuclear Physics, Polish Academy of Sciences*
*Kraków, Poland*
*E-mail:* oleh.kobzar@ifj.edu.pl

**Olexandr Sushchov**
*Chernihiv National Pedagogical University*
*Chernihiv, Ukraine*
*E-mail:* authule@yandex.ru

**Bohdan Hnatyk**
*Astronomical Observatory of Taras Shevchenko National University of Kyiv*
*Kyiv, Ukraine*
*E-mail:* bohdan_hnatyk@ukr.net

**Volodymyr Marchenko**
*Astronomical Observatory of the Jagellonian University*
*Kraków, Poland*
*E-mail:* volodymyr.marchenko@gmail.com

Active galactic nuclei (AGN) are considered as one of the most appropriate sources of cosmic rays with energy exceeding ~$10^{18}$ eV. Virgo A (M87 or NGC 4486) is the second closest to the Milky Way active galaxy. According to existing estimations it can be a prominent source of ultra high energy cosmic rays (UHECR). However not many events have been registered in the sky region near Virgo A, possibly due to magnetic field influence.

In the present work we check UHECR events from the recent sets of data (AUGER, Telescope Array etc.) for possibility of their origination in this AGN. We carried out the simulation of UHECR motion from Virgo A taking into account their deflections in galactic (GMF) as well as extragalactic (EGMF) magnetic fields according to the several latest models. The maps of expected UHECR arrival directions were obtained as a result.

It has been found the following: 1) UHECR deflection caused by EGMF is comparable with GMF one, moreover the influence of EGMF sometimes is dominating; 2) effect of EGMF demonstrates obvious asymmetry in the final distribution of expected UHECR arrival directions; 3) the results of simulations depend on chosen GMF model and are still open for the discussion.

[1]Speaker

## 1. Introduction

The origin of ultra high energy cosmic rays (UHECR) with energies exceeding ~$10^{18}$ eV, lying beyond the "ankle" in CR spectrum, is still the subject of discussions. It is known that they are represented mainly by protons and nuclei. Taking into account (1) the absence of appropriate sources inside our Galaxy and (2) almost uniform distribution of arrival directions [1 – 3] it is generally accepted that the origin of UHECR is extragalactic. Active galactic nuclei (AGN) as well as their jets and lobes are considered as one of the most appropriate sources of the cosmic rays with ultra high energy. This suggestion is confirmed by the presence of the concentrated group of UHECR events registered in the sky region close to Centaurus A (NGC 5128) [4, 5], which is the nearest AGN to Milky Way located at the distance of 3.8 Mpc. It was shown later, taking into account possible deflections of charged CR particles in the galactic magnetic field, that some of these events could really originate from Centaurus A [6, 7].

Virgo A (M87 or NGC 4486) is the second nearest AGN located at the distance of 16.4 Mpc. According to existing estimations it can be a prominent UHECR source as well. But in contrast to Centaurus A not many events have been registered near Virgo A. Moreover some zone of avoidance is observed in this sky region. Possibly it can be explained by significant influence of magnetic fields. Tacking into account larger distance to this AGN the influence of extragalactic field probably can not be neglected.

In the present work we carried out the simulation of UHECR motion from Virgo A taking into account their deflections in galactic (GMF) as well as extragalactic (EGMF) magnetic fields according to the several recent GMF models [8 – 12]. In result we have built the maps of expected UHECR arrival directions suggesting their different energies and particle types (protons and heavier nuclei). Also we compared obtained by different observatories (AUGER, Telescope Array etc.) maps of UHECR events detected during a long time of observations [1 – 4] and hence checked these events for possibility of their origin in Virgo A.

## 2. Method

### 2.1 Modelling of magnetic fields

Galactic magnetic field consists of two components: a regular and a random. We accent mainly upon the first one because its influence on the UHECR propagation is dominating. Random field can be taken into account as an amendment to the main effect of the regular component. Regular GMF in turn consists of a disk component and a field of galactic halo. The structure of the disk field is usually approximated by logarithmic spiral according to the matter distribution in the galactic plane. In halo field the toroidal and poloidal components are considered most often.

A lot of different GMF models are proposed at the present moment. We have chosen four most popular models developed during 2003 – 2012 [8 – 12]. The description of the disk field is quite similar in all considered models whereas differences for the halo field are more significant. The average GMF magnitude in all used models is of the order of several μG. It is important that all components of regular GMF can be described by empiric analytical formulas.

Extragalactic magnetic field has a random structure [13, 14]. It can be described in various ways, for example by the cubic cell model or by Kolmogorov spectrum modes. We used the model of cubic cells which is the simplest one. According to this model all space is divided on equal cells and MF is accepted to be uniform and only changes direction from cell to cell. Here we used the estimation of field magnitude ~1 nG with coherent length ~1 Mpc according to the criterion $\langle B\rangle\sqrt{l_0} \le 10^{-9}$ G·Mpc$^{-1/2}$ [13], where $l_0$ is the field coherent length.

### 2.2 Simulation of UHECR propagation

So far as regular GMF can be described explicitly, we calculated its influence on the UHECR trajectory by numerical solving of motion equations. In case of ultra relativistic particles with charge $q = Ze$ and energy $E \gg m_0 c^2$ it is useful to write these equations as $\frac{d\boldsymbol{r}}{d\,ct} = \hat{\boldsymbol{v}}$ & $\frac{d\hat{\boldsymbol{v}}}{d\,ct} = \frac{qc}{E}[\hat{\boldsymbol{v}} \times \boldsymbol{B}]$ , where $\hat{\boldsymbol{v}} = \frac{\boldsymbol{p}}{|\boldsymbol{p}|}$ is the unit vector of particle velocity.

The influence of EGMF can be described by CR diffusion due to its random structure. It results in scattering of arrival directions at the boundary of our Galaxy. This peculiarity allows us to calculate average UHECR deflection $\theta$ in EGMF statistically using the formula [15] $\langle\theta^2\rangle = \frac{2}{9}\left(\frac{qc}{E}\right)^2 \langle B^2\rangle\, l_0 L$ , where $L$ is the full distance covered by CR particle. Note, that scattering cone limited by the angle $\theta$ corresponds to the confidence interval $1\sigma$, containing 68% of total scattered particles.

The impact of random GMF component is similar to EGMF and was taken into account as a minor part of total scattering.

In our simulations we used back tracking method. Since among the registered UHECR the events which could be assigned to Virgo A are still absent, we have checked all directions in the northern galactic hemisphere with a step of 0.5 degrees. The motion of CR particles with different rigidity $E/Z$ was assumed. Within the Galaxy the simulated particles were moving under influence of only regular GMF, which was limited by the sphere of 50 kpc in radius. The EGMF was applied outside of this sphere as an overlay of scattering cone with angle $\theta$ upon outcoming direction at the edge of galactic sphere.

Finally we have chosen only those particles for which an outcoming direction together with the scattering cone traps the coordinates of Virgo A. Initial directions of these particles correspond to the expected UHECR arrival directions.

## 3. Results

As a result of carried out simulations we have built the maps of expected arrival directions for UHECR with rigidity $E/Z$ in the range $(\text{from } 5 \text{ to } 100) \times 10^{18}$ eV . It was realized by choosing the energy of simulated particles $E = 10^{20}$ eV and changing their $Z$ from 1 to 20. Note, that in general case these simulated values of $Z$ don't correspond to the real charge of particles and should be normalized depending of their energies.

All obtained results are shown on the following figures in galactic coordinates (fig. 1 – 4). Owing to large angular distance of Virgo A from galactic plane we demonstrate here only northern galactic hemisphere. The numbers around denote galactic longitude, latitude varies from 0° to 90° but it is not shown. Grid cells have size 15° × 15°.

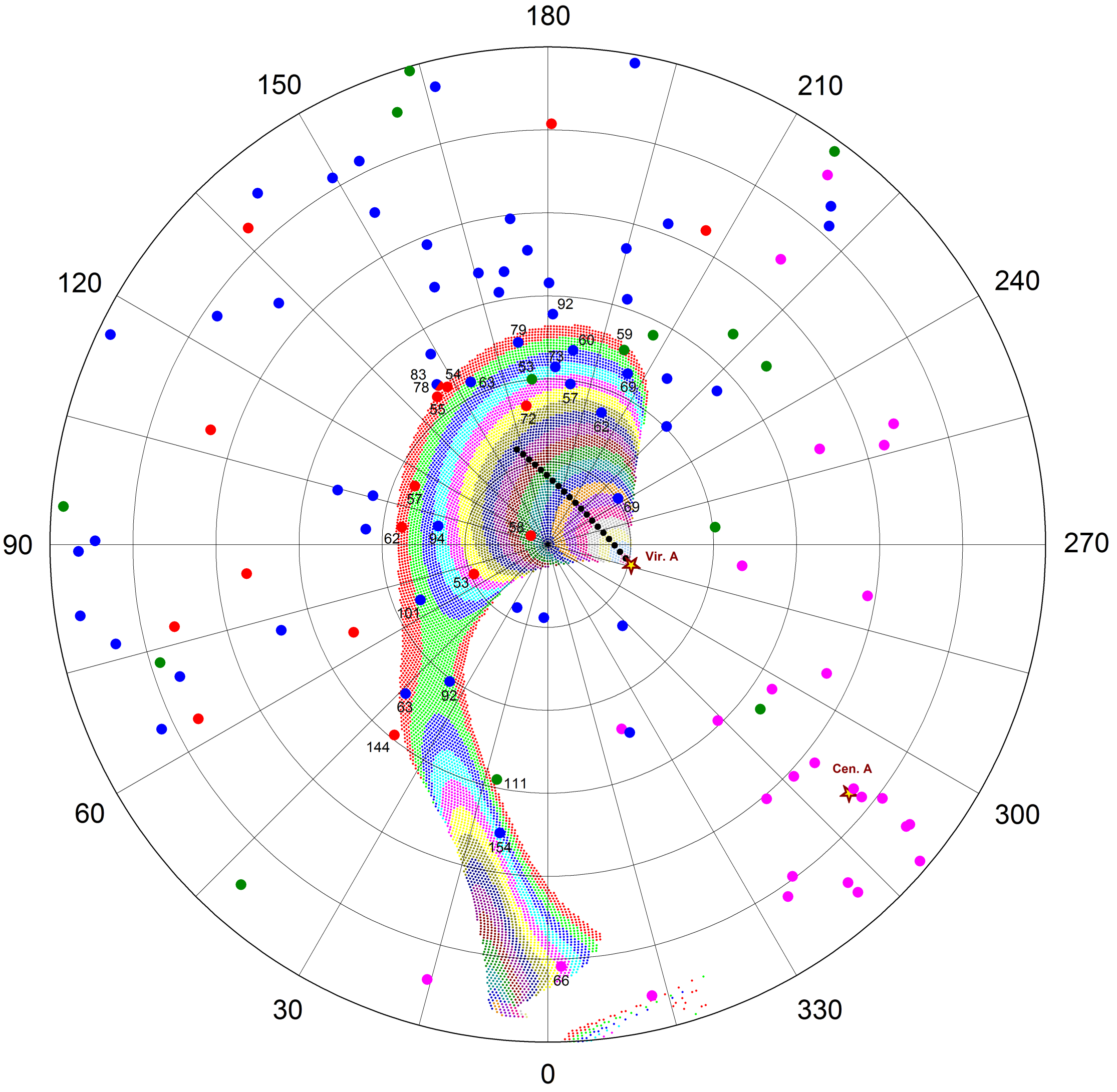

**Figure 1.** Map of expected arrival directions for UHECR from Virgo A based on GMF model published in 2003 [8].

AGN positions are denoted by stars. Black circles correspond to the expected events calculated for different $E/Z$ with taking into account only regular GMF component (the events with higher $Z$ are further from the AGN).

Fields of small coloured dots denote the regions of expected arrival directions expanded due to EGMF as well as random GMF influence. Corresponding colour denotation (the same for all figures 1 – 4) is following:

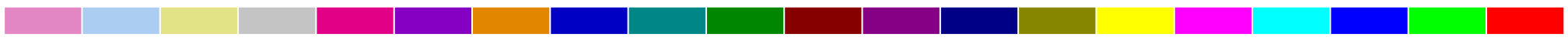

$\frac{E}{Z}$, EeV → $\frac{100}{1}$ $\frac{100}{2}$ $\frac{100}{3}$ $\frac{100}{4}$ $\frac{100}{5}$ $\frac{100}{6}$ $\frac{100}{7}$ $\frac{100}{8}$ $\frac{100}{9}$ $\frac{100}{10}$ $\frac{100}{11}$ $\frac{100}{12}$ $\frac{100}{13}$ $\frac{100}{14}$ $\frac{100}{15}$ $\frac{100}{16}$ $\frac{100}{17}$ $\frac{100}{18}$ $\frac{100}{19}$ $\frac{100}{20}$

Statistically these regions correspond to 68% of total CR amount with probability differing by factor not higher than 1.65.

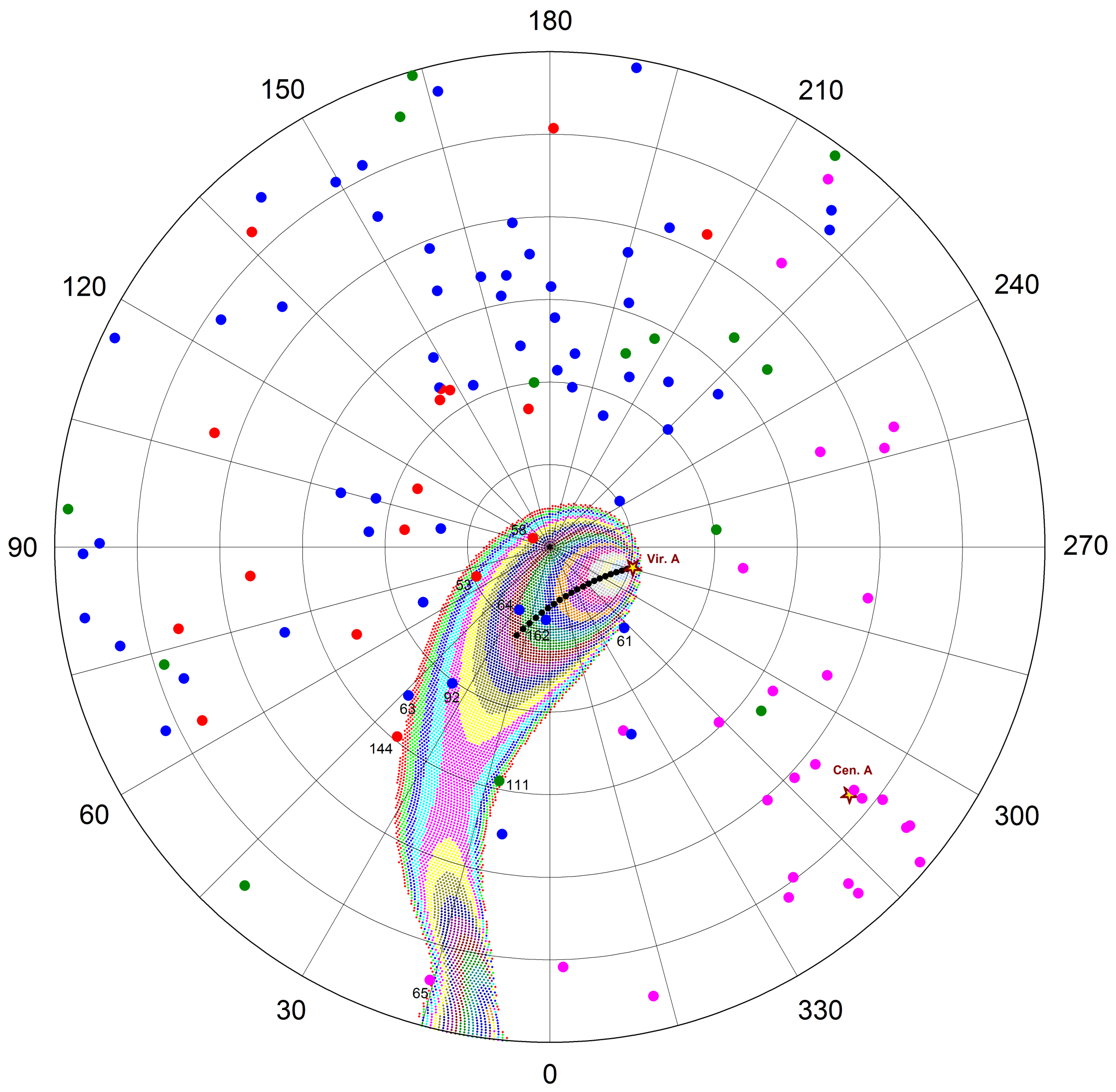

**Figure 2.** Map of expected arrival directions for UHECR from Virgo A based on GMF model published in 2007 [9].

Registered UHECR events are denoted by coloured circles: **AGASA** ($E > 50$ EeV) – red circles, **HiRes** ($E > 50$ EeV) – green circles, **AUGER** ($E > 55$ EeV) – magenta circles, **Telescope Array** ($E > 57$ EeV) – blue circles. Numbers near events denote corresponding energy in EeV ($10^{18}$ eV).

As one can see from presented results the influence of regular GMF on the UHECR deflection is strongly dependent on the chosen GMF model. At that time coupling with effect of EGMF all models demonstrate the similar trend to deflect UHECR arrival directions approximately towards the galactic centre. Also it seems that the influence of EGMF has the same order as GMF one and sometimes it even dominates. Moreover EGMF affects asymmetrically: the regions of expected UHECR arrival directions are obviously expanded in the direction perpendicular to the galactic plane.

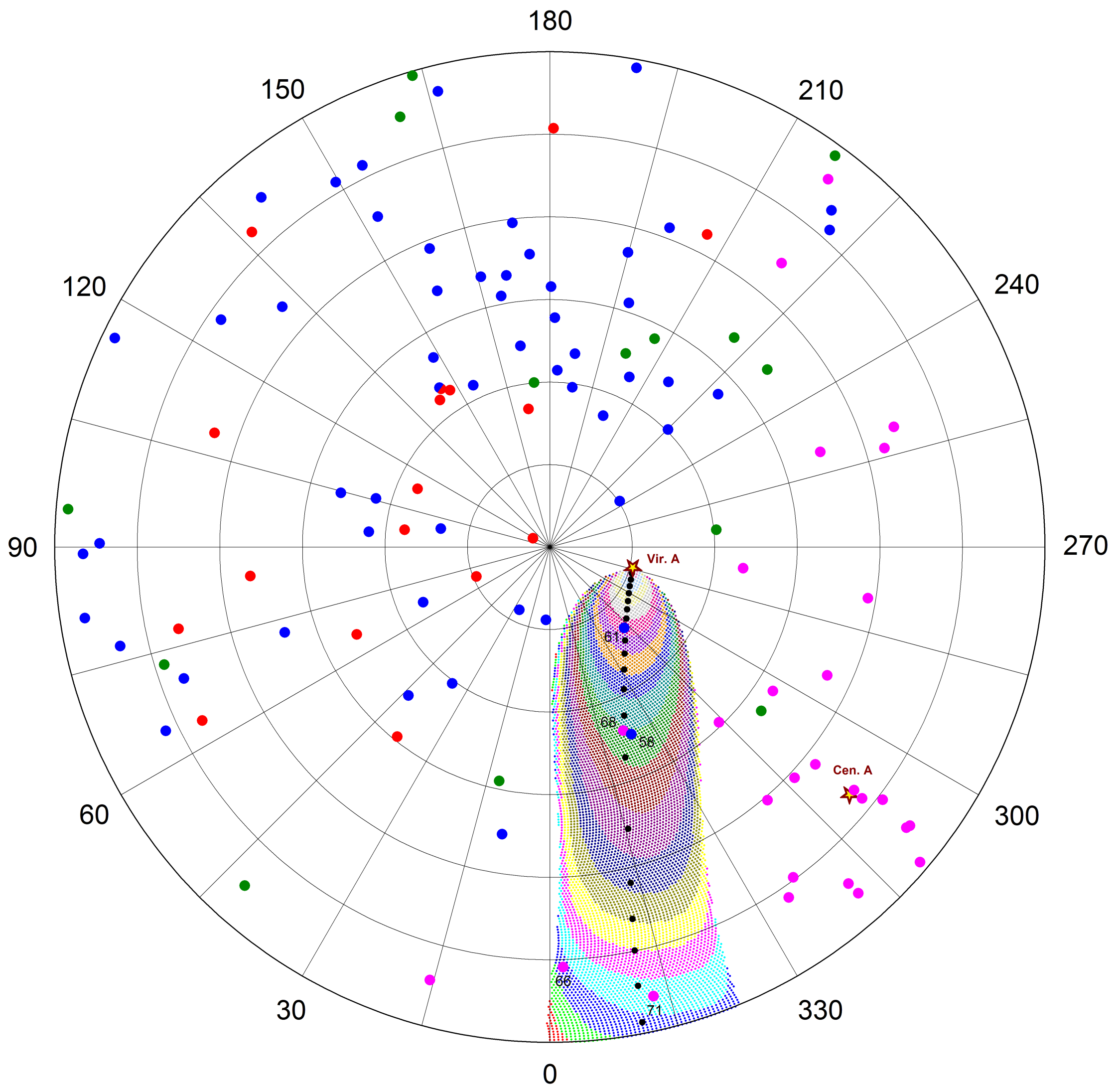

**Figure 3.** Map of expected arrival directions for UHECR from Virgo A based on GMF model published in 2011 [10].

Finally the use of different GMF models establishes correspondence of the various groups of events with Virgo A. Full sets of these events with energies and normalized values of $Z$ for each GMF model are listed in the table 1. Note, that each estimation of $Z$ corresponds to the bottom limit of this parameter, except only maximum values ~ 26 – 30. The highest value 30 corresponds to the event registered by AGASA and hence probably may be overestimated a little bit, mainly due to the common trend to slight overestimation of events energy by AGASA.

## 4. Conclusions

Deflection of UHECR from Virgo A caused by EGMF is generally comparable with GMF one, moreover the influence of EGMF can even dominate. Effect of EGMF demonstrates obvious asymmetry in the final distribution of expected UHECR arrival directions, including multi-images, as well as overlapping for different *Z*.

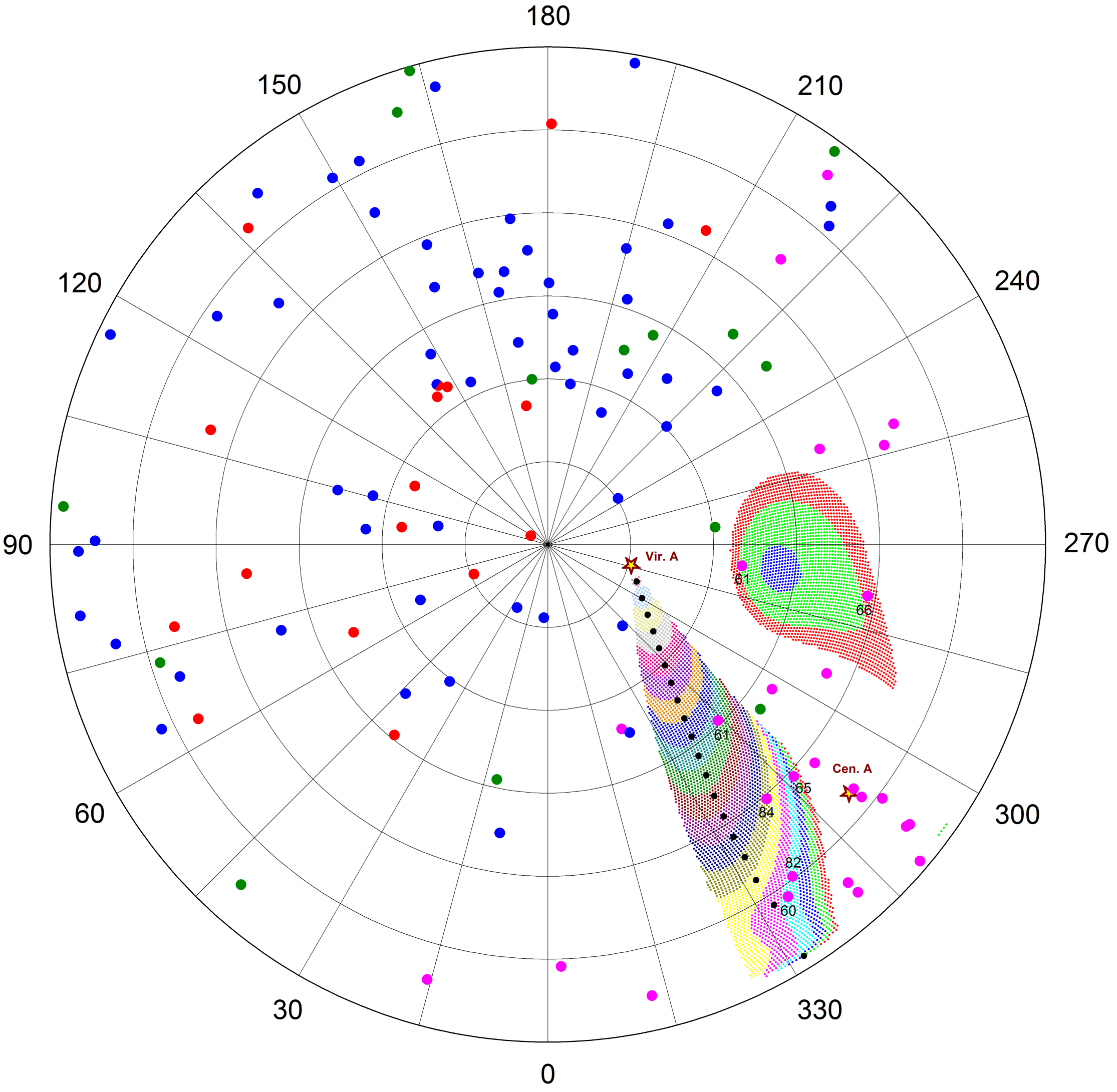

**Figure 4.** Map of expected arrival directions for UHECR from Virgo A based on GMF model published in 2012 [11, 12].

| model 2003 | | | | | | model 2007 | | model 2011 | | model 2012 | |
|---|---|---|---|---|---|---|---|---|---|---|---|
| *E*, EeV | *Z* | *E*, EeV | *Z* | *E*, EeV | *Z* | *E*, EeV | *Z* | *E*, EeV | *Z* | *E*, EeV | *Z* |
| 53 | 8 | 62 | 12 | 79 | 16 | 53 | 9 | 58 | 6 | 60 | 10 |
| 53 | 9 | 63 | 12 | 83 | 17 | 58 | 8 | 61 | 3 | 61 | 6 |
| 54 | 11 | 63 | 12 | 92 | 18 | 61 | 13 | 66 | 12 | 61 | 12 |
| 55 | 11 | 66 | 10 | 92 | 19 | 63 | 13 | 68 | 7 | 65 | 12 |
| 57 | 9 | 69 | 4 | 94 | 16 | 64 | 7 | 71 | 12 | 66 | 13 |
| 57 | 11 | 69 | 13 | 101 | 19 | 65 | 13 | | | 82 | 14 |
| 58 | 5 | 71 | 15 | 111 | 24 | 92 | 15 | | | 84 | 12 |
| 59 | 12 | 72 | 11 | 144 | 30 | 111 | 21 | | | | |
| 60 | 11 | 73 | 12 | 154 | 26 | 144 | 30 | | | | |
| 62 | 9 | 78 | 16 | | | 162 | 14 | | | | |

**Table 1.** List of UHECR events corresponding to Virgo A with use of different GMF models.

The results of our simulations allow to establish correspondence of the several groups of registered UHECR events with Virgo A. But the use of different GMF models establishes such correspondence for the various groups of events. They are following:
GMF model (2003) – 29 CR particles with $E$ from 53 to 154 EeV and $Z$ from 4 to 30.
GMF model (2007) – 10 CR particles with $E$ from 51 to 162 EeV and $Z$ from 7 to 30.
GMF model (2011) – 5 CR particles with $E$ from 58 to 71 EeV and $Z$ from 3 to 12.
GMF model (2012) – 7 CR particles with $E$ from 60 to 84 EeV and $Z$ from 6 to 14.
Despite of this fact all GMF models together with EGMF have a common trend to deflect UHECR from Virgo A towards the galactic centre.

## References

[1] N. Hayashida, K. Honda et al., *Updated AGASA event list above 4×10$^{19}$ eV*, arXiv: 0008102 [astro-ph]

[2] The Pierre Auger Collaboration, *Update on the correlation of the highest energy cosmic rays with nearby extragalactic matter*, ASTROPART PHYS **34** (2010) 314

[3] R. U. Abbasi, M. Abe et al., *Indications of Intermediate-Scale Anisotropy of Cosmic Rays with Energy Greater Than 57 EeV in the Northern Sky Measured with the Surface Detector of the Telescope Array Experiment*, ASTROPHYS J LETT **790** (2014) 006

[4] The Pierre Auger Collaboration, *Correlation of the highest-energy cosmic rays with the positions of nearby active galactic nuclei*, ASTROPART PHYS **29** (2008) 186

[5] D. Fargion, *UHECR besides Cen A: hints of galactic sources*, PROG PART NUCL PHYS **64** (2009) 363

[6] O. Sushchov, O. Kobzar, B. Hnatyk, V. Marchenko, *Search of ultra high cosmic rays' sources. FRI-radiogalaxy Centaurus A*, KINEMAT PHYS CELEST **28** (2012) 270

[7] G. R. Farrar, R. Jansson, I. J. Feain, B. M. Gaensler, *Galactic magnetic deflection and Centaurus A as a UHECR source*, J COSMOL ASTROPART P **01** (2013) 023

[8] M. Prouza, R. Šmida, *The Galactic Magnetic Field and Propagation of Ultra High Energy Cosmic Rays*, ASTRON ASTROPHYS **410** (2003) 001

[9] M. Kachelrieß, P. D. Serpico, M. Teshima, *The Galactic Magnetic Field as Spectrograph for Ultra High Energy Cosmic Rays*, ASTROPART PHYS **26** (2007) 378

[10] M. S. Pshirkov, P. G. Tinyakov, P. P. Kronberg, K. J. Newton-McGee, *Deriving Global Structure of the Galactic Magnetic Field from Faraday Rotation Measures of Extragalactic Sources*, ASTROPHYS J **738** (2011) 192

[11] R. Jansson, G. R. Farrar. *A New Model of the Galactic Magnetic Field*, ASTROPHYS J **757** (2012) 014

[12] R. Jansson, G. R. Farrar. *The Galactic Magnetic Field*, ASTROPHYS J LETT **761** (2012) L11

[13] P. P. Kronberg, *Extragalactic Magnetic Field*, REP PROG PHYS **57** (1994) 325

[14] D. Ryu, H. Kang, J. Cho, S. Das, *Turbulence and Magnetic Fields in the Large-Scale Structure of the Universe*, SCIENCE **320** (2008) 909

[15] V. S. Berezinsky, S. I. Grigorieva, B. I. Hnatyk, *Extragalactic UHE Proton Spectrum and Prediction for Iron-Nuclei Flux at 10$^{8}$ – 10$^{9}$ GeV*, ASTROPART PHYS **21** (2004) 617